\documentclass[12pt]{iopart}

\usepackage{amssymb}

\usepackage{epsf,graphicx,rotating,color}
\usepackage{bm,bbm}

\newcommand{\la}{\langle}
\newcommand{\ra}{\rangle}
\newcommand{\lla}{\left\langle}
\newcommand{\rra}{\right\rangle}

\newcommand{\eg}{\textit{e.g.} }

\newcommand{\fisguadalajara}{\address{Departamento de F\'isica, Universidad de 
  Guadalajara, Guadalajara, Jal., M\'exico}}

\begin{document}

\title[Scattering approach to fidelity decay]{Scattering approach to fidelity 
  decay in closed systems and parametric level correlations}

\author{T. Gorin and P. C. L\' opez V\' azquez}\fisguadalajara
 
\date{\today}

%

\begin{abstract}
This paper is based on recent work which provided an exact analytical 
description of scattering fidelity experiments with a microwave cavity under 
the variation of an antenna coupling [K\" ober et al., Phys. Rev. E 82, 036207 
(2010)]. It is shown that this description can also be used to predict the 
decay of the fidelity amplitude for arbitrary Hermitian perturbations of a 
closed system. Two applications are presented: First, the known result for 
global perturbations is re-derived, and second, the exact analytical expression 
for the perturbation due to a moving S-wave scatterer is worked out. The latter 
is compared to measured data from microwave experiments, which have been 
reported some time ago. Finally, we generalize an important relation between
fidelity decay and parametric level correlations to arbitrary perturbations.
\end{abstract}


\maketitle

\section{Introduction}

During the last decade, approximately, considerable efforts have been dedicated
to the quantitative prediction of the fidelity decay in chaotic/diffusive 
quantum systems and classical wave 
systems~\cite{ProZni01,CerTom02,VanHel03,Van05} (see also~\cite{GPSZ06} and 
references therein). A very successful approach has been based on random matrix 
theory, adopting the so called Bohigas-Giannoni-Schmit conjecture~\cite{BGS84}. 
Applied to the current setting, it suggests that quantum systems with chaotic 
classical counterpart (``chaotic quantum systems'' for short) as well as 
diffusive wave systems show a universal response to perturbations which can be 
calculated within an appropriate random matrix model~\cite{GPS04}. The first 
exact analytical results in this respect have been obtained by St\" ockmann and 
Sch\" afer~\cite{StoSch04b,StoSch05} using super-symmetry techniques similar
to those for the calculation of correlation functions between scattering matrix 
elements in~\cite{VWZ85}. More recently, exact analytical results have also 
been found for scattering systems, where the fidelity amplitude, an expectation 
value, is replaced by the ``scattering fidelity'', which is a product of two 
transition amplitudes~\cite{SGSS05,SSGS05}. These results, published 
in~\cite{KKSGSS10}, have been obtained by a simple but powerful modification of 
the Verbaarschot-Weidenm\" uller-Zirnbauer (``VWZ'' for short) formula 
from~\cite{VWZ85}. We call this approach the ``scattering approach to 
fidelity''. 

As shown first by Kohler et al.~\cite{Koh08}, for a global perturbation of a
completely diffusive system, the fidelity amplitude can also be calculated from 
the parametric level correlations. Subsequent generalizations have been 
discussed in~\cite{GutWal10,KoNaSt11} and~\cite{KohRec12}. Originally, 
parametric level 
correlations have been introduced in the area of disordered systems with 
diffusive dynamics~\cite{SzaAlt93,SimAlt93}. At that moment, they have been 
considered a universal signature of chaotic/diffusive dynamics, where the 
functional does not depend on the perturbation applied. However, 
in~\cite{BKS99} it was shown that certain types of perturbations may lead to 
pronounced deviations. Thereby, it became clear that the ``universal'' 
theoretical prediction of Simons and Altshuler~\cite{SimAlt93} only applies to 
global perturbations, not local ones, where the perturbation operator has only 
a few eigenstates with non-zero eigenvalues. The perturbation due to the 
displacement of a small scatterer discussed in~\cite{BKS99} is precisely of 
that latter type.

The purpose of the present paper is twofold. Firstly, we use the scattering 
fidelity approach from~\cite{KKSGSS10} to derive exact analytical expressions 
for the fidelity decay of chaotic/diffusive wave systems in the presence of 
completely general Hermitian perturbations. This allows us, to re-derive the 
known result for the decay of the fidelity amplitude due to a global 
perturbation~\cite{StoSch04b,StoSch05}. In the second example, we use it to 
derive an exact analytical expression for the decay of the fidelity amplitude 
due to the displacement of a S-wave scatterer (local perturbation). 
In~\cite{Hoe08}, the decay of the fidelity amplitude has been obtained from 
experimental for such a case~\cite{BKS99}. 

Secondly, we generalize the relation between the fidelity amplitude and the 
parametric level correlations from~\cite{Koh08}, to arbitrary perturbations. To 
do so, we compare the analytical expression for the parametric level 
correlations~\cite{MaSmSi03} and its analogue for the fidelity amplitude for 
general perturbations. Our result is important, as it allows to calculate the 
fidelity amplitude for spectral data, only. It thereby shows that the fidelity
amplitude is a basis independent -- which is surprising taking into account
that the perturbation may be completely arbitrary. From a practical point of
view, one may easily find situations, where the measurement of level spectra 
and their variation under certain perturbations is easier and more accurate 
than any fidelity measurement. The new relation shows, that a measurement of 
parametric level correlations provides exactly the same information about an a 
priori unknown perturbation than a fidelity measurement. 

The present paper is organized as follows: In the next section, we 
follow~\cite{KKSGSS10} to describe the connection between scattering 
fidelity~\cite{SGSS05} and scattering matrix correlation functions as 
considered in~\cite{VWZ85}. We use this connection to derive an exact 
analytical expression for the fidelity amplitude valid for arbitrary 
perturbations. In \sref{L} we discuss the differences between local and global 
perturbations, and we use our general formula to re-derive the known result for 
a global perturbation. In the remaining part of that section, we calculate the 
fidelity amplitude in the case of a moving scatterer and compare the resulting 
theoretical prediction to experimental data from~\cite{Hoe08}. In \sref{A}, we 
evaluate the general integral expression for the fidelity amplitude in the 
perturbative/long time limit. In \sref{P} we generalize the relation between 
parametric level correlations and the fidelity amplitude to arbitrary 
perturbations. Conclusions are presented in \sref{C}.

\section{\label{F} Scattering approach to fidelity}

In this section, we introduce the central quantity of this work, the fidelity 
amplitude of a closed quantum or classical wave system, with quantum chaotic or 
diffusive dynamics. We assume that random matrix theory can be used to describe 
the fidelity decay. While the first part contains some general statements about 
fidelity and the random matrix models used, the second part describes the 
description of the algebraic scattering model to which the fidelity problem is 
mapped. This mapping, introduced in~\cite{KKSGSS10} provides an exact 
analytical description of the fidelity decay.

\subsection{Fidelity}

The fidelity and the fidelity amplitude for a Hamiltonian 
$H_\alpha= H_0 + W_\alpha$ with perturbation $W_\alpha$ are defined as
\begin{equation}
F(t)= |f(t)|^2\; , \qquad
f(t) = \la a|\, \rme^{2\pi\rmi H_\beta\, t}\, \rme^{-2\pi\rmi H_\alpha\, t}\, 
  |a\ra
\label{fidstart}\end{equation}
where $|a\ra$ is the initial state and $W_\alpha$ is the perturbation depending
on an external parameter $\alpha$. We assume that the energy is measured in 
units of the mean level spacing $d_0$ in the spectrum of $H_0$, and time in 
units of the Heisenberg time $t_{\rm H} = 2\pi\hbar/d_0$. As a result, the 
variable $t$ in \eref{fidstart} becomes dimensionless.

For our purpose it will prove convenient to write the perturbation in terms of 
a normalized eigenbasis:
\begin{equation}
W_\alpha= \sum_j w_j(\alpha)\; |v_j\ra\la v_j| \; ,
\label{perturb}\end{equation}
where the orthonormal eigenstates $\{\, |v_j\ra \,\}$ are assumed to be
independent of $\alpha$. This is normally well fulfilled in the case of 
global perturbations, and also in the case of many types of local 
perturbations, such as point like scatterer. A detailed discussion is given in
\sref{L}. In other words, \eref{perturb} implies that $[W_\alpha,W_\beta] = 0$ 
for any $\alpha, \beta$ in the allowed range. Note that it is often possible to
consider $H_0 + W_\alpha$ as the unperturbed Hamiltonian $H_0'$. Then, 
$W_\alpha' = 0$ and $W_\beta'= W_\beta - W_\alpha$ such that 
$[W_\alpha',W_\beta']= 0$, trivially. 

Returing to our original setup, we choose $H_0$ from one of the invariant 
ensembles, the Gaussian orthogonal ensemble (GOE) or the Gaussian unitary 
ensemble (GUE)~\cite{Meh2004}. Correspondingly, we assume that the perturbation 
$W_\alpha$ can be diagonalized either by an orthogonal (GOE case) or unitary 
(GUE case) transformation. In either case, we arrive at
\begin{equation}
H_\alpha = H_0 + \sum_j w_j(\alpha)\; |j\ra\la j|\; ,
\label{F:Halpha}\end{equation}
without changes in the random matrix ensemble for $H_0$. Here, the states 
$|i\ra$ simply are the elements of the canonical basis of a complex vector 
space $\mathbb{C}^N$, where $N$ may be assumed arbitrarily large but finite. In 
this situation, one may use the results of~\cite{KKSGSS10} to calculate 
the fidelity amplitude averaged over $H_0$ as the average of a certain 
scattering matrix correlation function within the framework of statistical 
scattering~\cite{VWZ85}. In what follows, we concentrate on the GOE case. The 
GUE case (which turns out to be even simpler) may be treated along similar 
lines, using~\cite{SFS06}.

\subsection{Scattering matrix correlation functions}

According to~\cite{VWZ85}, the scattering matrix may be written as
\begin{equation}
S_{ab}(E)= \delta_{ab} - 2\rmi\pi\; V^\dagger\; \frac{1}{E - H_{\rm eff}}\; V
\; ,
\end{equation}
where $H_{\rm eff}= H_0 - \rmi\pi\; V\, V^\dagger$ with $H_0$ from the GOE. The 
indices $a,b$ denote the scattering channels which may be chosen in such a way 
that the column vectors of the rectangular matrix $V$ are orthogonal. Analogous 
to \eref{F:Halpha} it is thus possible to diagonalize the perturbation such 
that
\begin{equation}
H_{\rm eff}= H_0 - \rmi\pi\; \sum_a \gamma_a\; |a\ra\la a|
\label{Heff}\end{equation}
where the parameters $\gamma_a > 0$ are the real and positive eigenvalues of 
$V\, V^\dagger$ (eigenstates corresponding to zero eigenvalues are ignored). 
According to~\cite{VWZ85}, the average S-matrix (averaged over $H_0$), is
then given as
\begin{equation}
\mathbb{E}( S_{ab} ) = \frac{1 - \kappa_a}{1 + \kappa_a}\; , \qquad
\kappa_a = \frac{\pi^2\, \gamma_a}{N}\; ,
\label{couplps}\end{equation}
where we have assumed that the average level spacing for $H_0$ is equal to one. 
Here, we introduced the somewhat unusual notation $\mathbb{E}(\, \ldots\, )$
for the ensemble average over the Gaussian random matrix ensembles, to avoid
possible conflicts with the Bra-Ket notation used below. The main result 
of~\cite{VWZ85} consists in a triple integral which gives the spectral 
correlation function between different S-matrix elements 
\begin{equation}
\fl 
C[S_{ab}^*,S_{cd}](w) = \mathbb{E}\big [\, S_{ab}(E)^*\, S_{cd}(E+ wd)\, \big ]
 - \mathbb{E}[ S_{ab}(E)\ra^*]\; \mathbb{E}[ S_{cd}(E+ wd)] \; ,
\end{equation}
depending on the transmission coefficients
$T_a = 4\kappa_a\, (1+\kappa_a)^{-2}$, only. Due to the convolution theorem, 
the Fourier transform of these correlation functions yields an average over 
different amplitudes of the evolution operator for the effective Hamiltonian 
$H_{\rm eff}$. Namely, for $t>0$:
\begin{equation}
\hat C[S_{ab}^*,S_{cd}](t) \propto \la\hat S_{ab}(t)^*\, \hat S_{cd}(t)\ra
 = \mathbb{E}\big (\, \la b|\rme^{2\pi\rmi H_{\rm eff}^\dagger\, t} |a\ra
   \la c|\rme^{-2\pi\rmi H_{\rm eff}\, t} |d\ra\, \big ) \; , 
\label{corrfun}\end{equation}
It is still assumed that the Hamiltonian is written in the eigenbasis of the 
coupling $VV^\dagger$. Therefore, the states $|a\ra, |b\ra, |c\ra$ and $|d\ra$ 
represent elements of the canonical basis in $\mathbb{C}^N$. In what follows, 
we will only be concerned with the case when $c=a$ and $d=b$. This yields for
the correlation function in~\eref{corrfun}:
\begin{equation}
\hat C[S_{ab}^*,S_{ab}](t) = \delta_{ab}\, T_a^2\, \la Z\, J_a^2\ra_I 
   + (1+\delta_{ab})\, T_a T_b\, \la Z\, P_{ab}\ra_I \; ,
\label{F:VWZresult}\end{equation}
where the angular brackets $\la\ldots\ra_I$ denote the following 
weighted double-integral:
\begin{equation}
\la\ldots\ra_I = \int_{{\rm max}(0,t-1)}^t\rmd r\int_0^r\rmd u\, 
  \frac{(t-r)(r+1-t)}{(2u+1)(t^2-r^2+x^2)^2} \; ,
\label{cfundoubleint}\end{equation}
and where we have introduced the following short hands:
\begin{equation}
Z= \prod_{j=1} \frac{1-T_j(t-r)}{\sqrt{1+2T_j\, r + T_j^2\, x^2}}\; , \quad
x^2 = u^2\, \frac{2r+1}{2u+1} \; .
\label{defofZ}\end{equation}
Note that changing the integration variable from $u$ to $x$ yields
\begin{equation}
\fl 2x\, \rmd x = (2r+1) \left[ \frac{2u}{2u+1} - \frac{2u^2}{(2u+1)^2}\right] 
\rmd u \quad\Rightarrow\quad
\frac{\rmd u}{2u+1} = \frac{\rmd x}{\sqrt{x^2 + 2r+1}} \; ,
\end{equation}
such that \eref{cfundoubleint} may be written equivalently as
\begin{equation}
\la\ldots\ra_I = \int_{{\rm max}(0,t-1)}^t\rmd r\int_0^r\rmd x\, 
  \frac{(t-r)(r+1-t)}{\sqrt{x^2+2r+1}\, (t^2-r^2+x^2)^2} \; .
\label{cfundoubleint2}\end{equation}
That expression can be compared directly to the results 
in~\cite{StoSch04b,StoSch05}.

\subsection{\label{FC} Connection to fidelity}

Starting from~\eref{fidstart}, we insert the projection onto a random state 
$|b\ra\la b|$ into the definition of the fidelity amplitude:
\begin{equation}
f(t) \to f_{ab}(t) \propto \mathbb{E}\big (\, \la a|
  \rme^{2\pi\rmi H_\beta\, t}\, |b\ra\la b|\, \rme^{-2\pi\rmi H_\alpha\, t}\, 
  |a\ra \, \big ) \; ,
\label{F:scattfid}\end{equation}
where the average over the random state $|b\ra\la b|$ simply yields the 
identity times a normalization constant equal to the inverse Hilbert 
space dimension. As a result, we obtain the product of two transition 
amplitudes which may be considered as a scattering fidelity as introduced 
in~\cite{SGSS05}. Comparing the effective Hamiltonian~\eref{Heff} with the
perturbed Hamiltonian for a closed system as given in~\eref{F:Halpha}, we find 
that they share the same structure, and that we only need to allow the coupling 
parameters $\gamma_a$ to become complex to unify both descriptions.

In~\cite{KKSGSS10}, it has then been shown, that the analytical result for the 
correlation functions in~\eref{F:VWZresult} can be generalized to different
effective Hamiltonians $H_{\rm eff}$ and $H_{\rm eff}'$, which differ only in 
the eigenvalues $\gamma_a$ and $\gamma_a'$. In that case, one just needs to 
calculate the effective transmission coefficients
\begin{equation}
T_j = \frac{2\, (\kappa'_j + \kappa_j^*)}{(1+\kappa'_j)(1+\kappa_j^*)} 
\label{efftramicos}\end{equation}
from the coupling parameters $\kappa_a$ (corresponding to $H_{\rm eff}$) and 
$\kappa_a'$ (corresponding to $H_{\rm eff}'$), as defined in~\eref{couplps}. 
Then, the double integral in~\eref{cfundoubleint} yields the scattering 
fidelity, when replacing the transmission coefficients in the term $Z$ by the 
effective transmission coefficients defined in~\eref{efftramicos}. 

Restricting ourselves to closed systems with Hermitian perturbations, we obtain 
from the comparison of~\eref{F:Halpha} with~\eref{Heff} that 
$-\rmi\pi\gamma_j = w_j(\alpha)$, such that according to~\eref{couplps}
\begin{equation}
\kappa_j = \frac{\pi^2\, \gamma_j}{N} = \frac{\rmi\, \pi\, w_j(\alpha)}{N}
\; , \qquad
\kappa_j' = \frac{\rmi\, \pi\, w_j(\beta)}{N} \; .
\label{FC:couplpars}\end{equation}
Finally, to make sure that we really have a closed systemxs, we need the 
transmission coefficients $T_a$ and $T_b$ to be negligibly small. This means 
that the dynamics of the system is probed from the outside via scattering 
channels which are so weakly coupled to the system, that their effect on the 
dynamics is negligible. The functions to be integrated in \eref{F:VWZresult}
then become
\begin{eqnarray}
J_a &\to& 2t \nonumber\\
P_{ab} &\to& P_0 =  2\, \big [\, r^2 + (2r+1)\, t -t^2 - x^2\, \big ] \; .
\label{JPlim}\end{eqnarray}
Thereby, we obtain for the scattering fidelity
\begin{equation}
f_{ab}(\{\kappa_j\},\{\kappa_j'\}\, ;\, t) \propto 
  \delta_{ab}\; T_a^2\; 4t^2\; \la Z\ra_I + (1 + \delta_{ab})\;
  T_a\, T_b\; \la Z\; P_0\ra_I \; .
\end{equation}
Here, we indicate explicitly the dependence of the scattering fidelity on 
the coupling parameters $\{\kappa_j\}$ and $\{\kappa_j'\}$ as their value
will become important below, where we discuss normalization.

\subsubsection*{Normalization}

In order to calculate the fidelity amplitude from the scattering fidelity
$f_{ab}(t)$, there is still the problem of normalization to be solved.
This is because $f_{ab}(t)$ becomes an auto correlation function for
$H_{\rm eff} = H_{\rm eff}'$, which still decays to zero in time, if the 
coupling to decay channels is finite. In~\cite{SGSS05}, this problem has
been solved by dividing the scattering fidelity through the geometric mean of 
the auto correlation functions of $H_{\rm eff}$ and $H_{\rm eff}'$. Below, we
will see that this normalization procedure is somewhat simpler in the case of
closed systems.

As mentioned earlier, when one is really interested in fidelity and 
$|b\ra\la b|$ has been inserted just for convenience as described in
\eref{F:scattfid}, one can normally assume that $a\ne b$. In addition, the 
case $a\ne b$ arises in the case of an explicit scattering fidelity 
experiment, where in- and out-going channels are chosen to be different 
(transmission measurement). Then, the formula for $f_{ab}(t)$ simplifies to
\begin{equation}
f_{ab}(t) \propto  T_a T_b\; \la P_0\; Z\ra_I \; .
\end{equation}
In order to apply the normalization scheme from~\cite{SGSS05}, we note 
that for the auto correlation functions:
\begin{equation}
f_{ab}(\{\kappa_j\},\{\kappa_j\}\, ;\, t) 
 = f_{ab}(\{\kappa_j'\},\{\kappa_j'\}\, ;\, t) \propto T_a T_b\; \la P_0\ra_I
\; .
\end{equation}
This follows from the fact that 
$\kappa_j + \kappa_j^* = \kappa_j' + \kappa_j'{}^* = 0$ since the coupling
parameters are purely imaginary in both cases. That implies that the effective
transmission coefficients are zero, so that $Z$ becomes equal to one. Since
the auto correlation functions are the same, the geometric mean is also the
same, and
\begin{equation}
f_{ab}(\{\kappa_j\},\{\kappa_j'\}\, ;\, t) 
  = \frac{T_a T_b\; \la P_0\; Z\ra_I}{ T_a T_b\; \la P_0\ra_I} 
  = \frac{\la P_0\; Z\ra_I}{\la P_0\ra_I} \; .
\end{equation}
Now, it has been shown in~\cite{StoSch04b} that for $Z=1$, the resulting 
double integral yields
\begin{equation}
\la P_0\ra_I = \int_{{\rm max}(0,t-1)}^t\rmd r\int_0^r\rmd x\, 
  \frac{(t-r)(r+1-t)\, P_0}{\sqrt{x^2+2r+1}\, (t^2-r^2+x^2)^2} = 1
\end{equation}
for any $t > 0$, so that
\begin{equation}
f_{ab}(\{\kappa_j\},\{\kappa_j'\}\, ;\, t) = \la P_0\; Z\ra_I \; .
\label{F:fab}\end{equation}
This formula constitutes the first important result of our work, since it
gives an exact analytical expression for the fidelity amplitude of a 
chaotic/diffusive wave system for an arbitrary perturbation.

In the special case, when the scattering fidelity is measured from a reflection
amplitude ($a=b$), we find
\begin{equation}
f_{aa}(\{\kappa_j\},\{\kappa_j'\}\, ;\, t)= N(t)^{-1}\; T_a^2\; 
  \big [\, 4t^2\; \la Z\ra_I + 2\; \la Z\; P_0\ra_I\, \big ] \; ,
\end{equation}
where the geometric mean of the auto correlation functions, denoted by $N(t)$
turns out to be time dependent. While the effective transmission coefficients
are still zero and $Z=1$, the auto correlation functions now read:
\begin{equation}
\fl f_{aa}(\{\kappa_j\},\{\kappa_j\}\, ;\, t)
 = f_{aa}(\{\kappa_j'\},\{\kappa_j'\}\, ;\, t) = N(t)
 = T_a^2\; \big [\, 4t^2\; \la 1\ra_I + 2\; \la P_0\ra_I\, \big ] \; .
\end{equation}
The integral $\la 1\ra_I$ has been calculated in~\cite{GS02}, with the
result: $4 t^2\; \la 1\ra_I = 1 - b_2(t)$, where $b_2(t)$ is the two-point 
spectral form factor of the Gaussian orthogonal ensemble~\cite{Meh2004}.
Hence, we obtain:
\begin{equation}
f_{aa}(\{\kappa_j\},\{\kappa_j'\}\, ;\, t) 
 = \frac{4t^2\; \la Z\ra_I + 2\; \la Z\; P_0\ra_I}{3 - b_2(t)} \; .
\label{F:faa}\end{equation}
This result will be used in \sref{LL}, where we discuss experimental results
for perturbations due to the displacement of an S-wave scatterer.

\section{\label{L} Local vs. global perturbations}

A detailed discussion of the differences between local and global 
perturbations can be found in~\cite{MaSmSi03}. Consider~\eref{perturb}, where a 
perturbation results in the change of several eigenvalues of the perturbation 
operator $W$. In order to affect the dynamics of the system (leading to the 
decay of the fidelity amplitude), one may either change only a few eigenvalues 
by a large amount (local perturbation) or very many eigenvalues by a small 
amount (global perturbation). Both cases are considered in this section.

\subsection{\label{LG} Global perturbation}

This was the first problem solved in the context of fidelity decay of 
quantum-chaotic systems~\cite{StoSch04b,StoSch05,SSGS05}. Experimentally, the 
perturbation was realized in a chaotic microwave billiard by displacing one 
of the straight billiard boundaries. If described by~\eref{fidstart} 
and~\eref{perturb}, $W_\alpha$ may represent absolute displacements with 
respect to some initial position. Then, its eigenvector representation
\begin{equation}
W_\alpha = \sum_{j=1}^N w_j(\alpha)\; |v_j\ra\la v_j|
\end{equation}
runs over a large number $N$ of states. According to \sref{FC}, and in 
particular \eref{efftramicos} and \eref{FC:couplpars}, the effective 
transmission coefficients become
\begin{eqnarray}
T_j &=& \frac{2\pi}{N}\; \frac{\rmi\, w_j(\beta) - \rmi\, w_j(\alpha)}
  {[1 + \rmi\pi\, w_j(\beta)/N]\, [1 - \rmi\pi\, w_j(\alpha)/N]} \nonumber\\
 &=& 2\pi\rmi\, \delta_j\, (1 - \rmi\pi\, \delta_j) + \mathcal{O}\Big (\,
   [w_j(\beta)/N]^3\, ,\, [w_j(\alpha)/N]^3\, \Big ) \; ,
\label{LG:Teffkapp}\end{eqnarray}
where $\delta_j= [w_j(\beta) - w_j(\alpha)]/N$. In this setting, global 
perturbations are characterized by the fact that the contribution of each 
individual term is negligible, while the perturbation becomes noticeable only 
because it is the sum of very many such contributions. This allows to perform a 
Taylor expansion of $\ln Z$ with respect to the coupling parameters $\delta_j$. 
Starting from the Taylor expansion of $Z_j$ with respect to the transmission 
coefficients
\begin{eqnarray}
Z_j &= [1 - T_j\, (t-r)]\; [1 + 2 T_j\, r + T_j^2\, x^2]^{-1/2}\nonumber\\
    &= [1 - T_j (t-r)]\; [1 - r T_j - (x^2-3r^2)\, T_j^2/2] 
  + \mathcal{O}(T_j^3) \nonumber\\
 &= 1 - t\; T_j + [rt +r^2/2 - x^2/2 ]\; T_j^2 + \mathcal{O}(T_j^3) \; , 
\end{eqnarray}
we insert \eref{LG:Teffkapp} into the above Taylor expansion, and obtain 
\begin{eqnarray}
\fl \ln Z = -2\pi\rmi\sum_j \delta_j\, t 
  - 2\pi^2\sum_j \big (\, r^2 + (2r+1)\, t -t^2 - x^2\, \big )\; \delta_j^2
\nonumber\\
 + \mathcal{O}\Big (\, [w_j(\beta)/N]^3\, ,\, [w_j(\alpha)/N]^3\, \Big ) \; .
\end{eqnarray}
In order to obtain a well defined function $Z(t,r,x)$ in the limit of
$N\to\infty$, and vanishing perturbation: $\delta_j\to 0$, the parameters
$\delta_j$ must scale with an appropriate negative power of $N$: (i) For 
$\delta_j \sim N^{-1}$, $\ln Z$ would converge to the finite value 
$-2\pi\rmi \sum_j \delta_j\, t$. (ii) For $\delta_j \sim N^{-1/2}$, the sum 
$\sum_j \delta_j$ could still converge, if the $\delta_j$ had different signs. 
In addition, the sum $\sum_j \delta_j{}^2$ would always converge, while any
higher order terms would vanish. (iii) For powers larger than $-1/2$, the sum 
$\sum_j \delta_j{}^2$ would always diverge, and the function $Z(t,r,x)$ would 
not be well defined. Hence, the cases (i) and (ii) are the only viable 
options, where
\begin{equation}
\sum_j \delta_j = \delta_{\rm s}\; , \qquad
\sum_j \delta_j{}^2 = \lambda^2
\label{LG:kapisums}\end{equation}
converge to finite values. This finally leads to 
\begin{equation}
\lim_{N\to\infty} Z = \exp( -2\pi\rmi\, \delta_{\rm s}\, t 
   - \pi^2\lambda^2\, P_0 ) \; ,
\label{LG:Z}\end{equation}
with $P_0$ given in~\eref{JPlim}. Note that by taking the absolute value
squared of the fidelity amplitude, the dependence on $\delta_{\rm s}$
disappears and with it any possible problems with the convergence of this
term.

To conclude this section about global perturbations, let us discuss the random 
matrix model for fidelity decay, as it has been first introduced 
in~\cite{GPS04}. This model may be written as
\begin{equation}
H_\alpha = H_0 + \alpha\; V\; , 
\end{equation}
where the matrices $H_0$ and $V$ are independent GOE matrices, 
normalized in such a way that the mean level spacing in the centre of the 
spectrum of $H_0$ is $d_0=1$, while for the perturbation matrix it holds
\begin{equation}
\la V_{ij}\, V_{kl}\ra = \delta_{jk}\delta_{il} + \delta_{ik}\delta_{jl} \; .
\end{equation}
Now, representing $H_\alpha$ in the eigenbasis of $V$, the perturbation 
becomes diagonal with eigenvalues $w_j(\alpha)$ showing a semi-circle 
distribution between $-2\alpha\sqrt{N}$ and $2\alpha\sqrt{N}$. Then, according
to~\eref{perturb}:
\begin{equation}
\kappa_j = 0\; , \qquad \kappa_j' = \frac{\rmi\pi}{N}\; w_j(\alpha) 
\quad\Rightarrow\quad \delta_j= \frac{w_j(\alpha)}{N} \; ,
\end{equation}
which is of order $N^{-1/2}$, indeed. From the semi-circle distribution of the
eigenvalues $w_j(\alpha)$ it follows that
\begin{equation}
\sum_j \delta_j = \frac{1}{N}\sum_j w_j(\alpha)
  = 0 \; , \qquad
\sum_j \delta_j{}^2 = \frac{1}{N^2}\; \sum_j w_j(\alpha)^2
  = \alpha^2 \; .
\end{equation}
This shows that~\eref{LG:Z} applies for this case if we set 
$\delta_{\rm s} = 0$ and $\lambda = \alpha$. From~\eref{F:fab} it then follows 
that 
\begin{equation}
f_{ab}(t)= \lla P_0\; \rme^{-\pi^2\lambda^2\, P_0}\rra_I \; ,
\label{FidGlobal}\end{equation}
which agrees precisely with the result obtained in~\cite{StoSch04b}.

\subsection{\label{LL} Local perturbations}

In the case of local perturbations, $W_\alpha$ and $W_\beta$ differ strongly in
a relatively small subspace. A Taylor series expansion as in the previous case
is therefore not useful, and it is also less likely that 
$[W_\alpha,W_\beta] =0$. Thus, it seems necessary to redefine the perturbation 
by considering $H_\alpha$ as the unperturbed system and $W_\beta - W_\alpha$ as 
the perturbation. In doing so, it is assumed that including $W_\alpha$ into 
$H_0$ doesn't change its statistical (i.e. random matrix) properties. We may 
then choose a basis in which $W_\beta - W_\alpha$ is diagonal, the 
transformation into that basis leaves the random matrix ensemble for the new 
$H_0$ invariant, so that we arrive again at a description in which the 
perturbation is diagonal. 

In contrast to the previous case, we have now only a small number of non-zero 
diagonal elements, while each of them may be very large. In principle, each 
element alone can cause the fidelity to decay as fast as in the case of a 
global perturbation. In what follows, we restrict ourselves to the least biased 
case, where the states $|a\ra$ and $|b\ra$ coupled to the measurement channels 
[cf.~\eref{corrfun} and~\eref{F:VWZresult}] are not involved into the 
perturbation. In this case, the results become again independent from the 
measurement channels, and one can then repeat the previous argument to show 
that scattering fidelity (if measured in transmission) and fidelity amplitude 
must coincide.

\subsubsection{Moving scatterer}

This case refers to the displacement of a small scatterer from a position 
$\vec r_1$ to another position $\vec r_2$. As explained above, this is modelled 
by the unperturbed Hamiltonian consisting of the system with scatterer at 
position $\vec r_1$, while the perturbation consists in removing the scatterer 
from position $\vec r_1$ and placing it at position $\vec r_2$. For a 
point-like scatterer, the effect of the scatterer can be described by one 
single state (the perturbation operator corresponding to that scatterer has 
only one non-zero eigenvalue). Therefore,
\begin{equation}
H_\alpha = H_0\; , \qquad H_\beta = H_0 + w(\beta)\; \big (\,
    |v_2\ra\la v_2| - |v_1\ra\la v_1|\, \big ) \; .
\end{equation}
For the scattering approach to fidelity, this means that the perturbation must
be described by two effective transmission coefficients:
\begin{equation}
T_1 = \frac{-2\pi\rmi\, \delta_1}{1-\rmi\pi\, \delta_1}\; , 
\quad T_2 = \frac{2\pi\rmi\, \delta_1}{1+\rmi\pi\, \delta_1}\; , \qquad 
\delta_1 = \frac{w(\beta)}{N} \; .
\label{LL:T1T2}\end{equation}
Here, $\delta_1$ may become arbitrarily large so that $T_1$ as a function
of $\delta_1$ traces a path in the complex plane which starts at $T_1= 0$
and ends at $T_1 = 2$, while $T_2 = T_1^*$. Then
\begin{equation}
Z= \prod_{j=1}^M \frac{1- T_j(t-r)}{\sqrt{1+ 2T_j r + T_j^2 x^2}} =
\frac{|1 - T_1\, (t-r)|^2}{|1+2T_1\, r + T_1^2 x^2|} 
\end{equation}
inserted into \eref{F:fab} or \eref{F:faa} yields the exact analytical 
expression for the decay of the fidelity amplitude.

\subsubsection{Comparison with experiment}

The perturbation we just described, applies precisely to an experiment 
published in~\cite{BKS99,Hoe08}. There, a small disk of diameter
$4.6\, {\rm mm}$ has been moved in steps of $|\Delta r|= 1\, {\rm mm}$ through
a rectangular two-dimensional microwave billiard with 19 additional random
scatterers. Then, the reflection spectrum has been measured for 300 different 
positions of the moving disk in a frequency range from $3.5$ to 
$6\, {\rm GHz}$. In this frequency range, it was still possible to extract 
resonance positions and amplitudes by Lorentzian fits. The statistical 
properties of the spectrum as well as the wave functions were in agreement with 
the random matrix expectation for quantum chaotic or weakly disordered systems.

\begin{figure}
\includegraphics[width=0.8\textwidth]{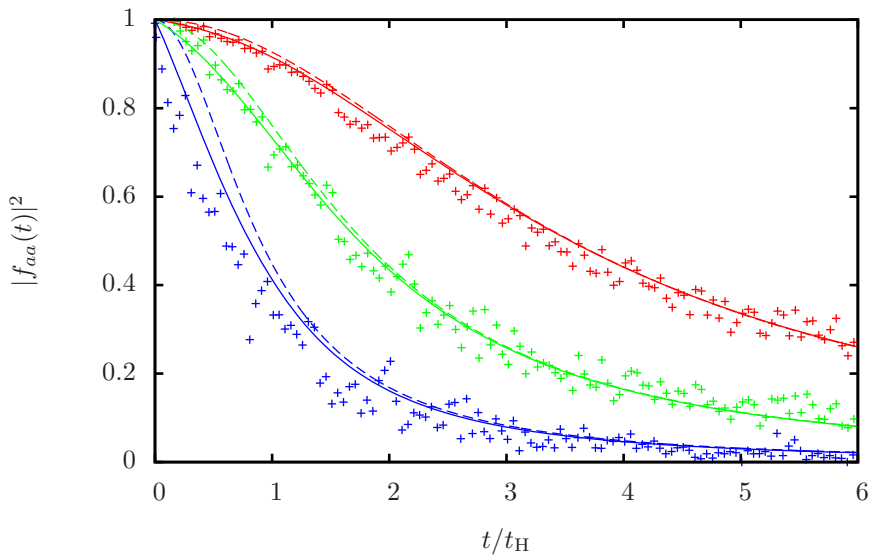}
\caption{Experimental data for the fidelity decay due to a moving scatterer
from~\cite{Hoe08}, compared to the approximate (dashed lines) and to the 
exact theory (solid lines). The different colours, red, green, and blue, 
correspond to different displacements $\Delta r = 1\, {\rm mm}$ 
($\delta_1 \approx 0.07$), $2\, {\rm mm}$ ($\delta_1 \approx 0.14$), and 
$4\, {\rm mm}$ ($\delta_1 \approx 0.28$), respectively.}
\label{Fig2}\end{figure}

From Berry's model of the random superposition of plane waves~\cite{Berry77}, 
it is possible to obtain a connection between the displacement $|\Delta r|$ of 
the movable disk and the parameter $\delta_1$ which measures the perturbation 
strength. Translating the corresponding equation from~\cite{Hoe08} to our
system of units and parameters, we obtain
\begin{equation}
\delta_1 = \frac{\alpha}{4}\; \sqrt{1 - J_0(k |\Delta r|)^2} \; , \qquad
\label{BRW:Eq1}\end{equation}
where $\alpha$ is a dimensionless factor related to the electromagnetic
properties of the movable disk, which can be determined independently 
({\it e.g.} from the variance of the level velocities). For the wavenumber $k$,
we choose a value which corrresonds to the frequency in the centre of the range 
considered, which yields
\begin{equation}
k= \frac{2\pi\, f}{c} = 0.996\, {\rm cm}^{-1} \; .
\end{equation}
For the displacements considered in~\cite{Hoe08}, the relation between 
$\delta_1$ and $|\Delta r|$ is still approximately linear, as can be seen from 
the fact that $k\, |\Delta r|$ is small as compared to one in all cases (see 
the captions of \fref{Fig2}). Finally, we find that $\alpha = 1$ provides the 
best agreement between the theory and experiment. Using time independent 
perturbation theory, the authors of~\cite{Hoe08} obtained for the decay of the 
fidelity amplitude the following asymptotic result:
\begin{equation}
f(t)= \frac{1}{\sqrt{1 + (4\delta_1 t)^2}} \; , 
\label{BRW:Eq2}\end{equation}
valid for finite $\delta_1\, t$, in the case where $\delta_1 \to 0$ and 
$t\to\infty$, and in agreement with our asymptotic result~(\ref{A:movscat}),
derived below (\sref{A}).

In \fref{Fig2} we show the experimental data for the decay of the absolute
value squared $|f_{aa}(t)|^2$ as obtained in~\cite{Hoe08}. In the
experiment, this quantity is obtained from ensemble averages of the respective
correlation functions, introduced in~\eref{F:scattfid}. The results are
compared to the theoretical predictions based on the perturbative 
result,~\eref{BRW:Eq2}, and on our exact analytical expression for a 
reflection measurement,~\eref{F:faa}. 

We focus here on the behaviour of the fidelity at small times, where the 
asymptotic formula is expected to be less accurate, and indeed, we find 
significant deviations for the cases $\delta_1\approx 0.14$ (green points vs. 
dashed green line) and $\delta_1\approx 0.28$ (blue points vs. dashed blue 
line). For these cases, the experimental fidelity decay has a notable linear 
component at small times, which cannot be reproduced by the perturbative 
formula,~\eref{BRW:Eq2}. By contrast, our exact analytical result contains that 
linear component and therefore agrees much better with the experiment (solid 
lines). Still, some differences remain for $\delta_1\approx 0.28$. We believe 
that these are due to problems on the experimental side. One error source 
consists in the wide frequency range used, which, according to~\eref{BRW:Eq1} 
leads to a considerable variation in the perturbation strength. Another problem 
is related to the upper end of the frequency range, which implies rather small 
wavelengths, for which the scatterer to be moved may no longer be point like. 
For a more significant test of our analytical formula, one would need a 
different experimental design, providing higher accuracies at strong
perturbations in the vicinity of the Heisenberg time.

\begin{figure}
\includegraphics[width=0.9\textwidth]{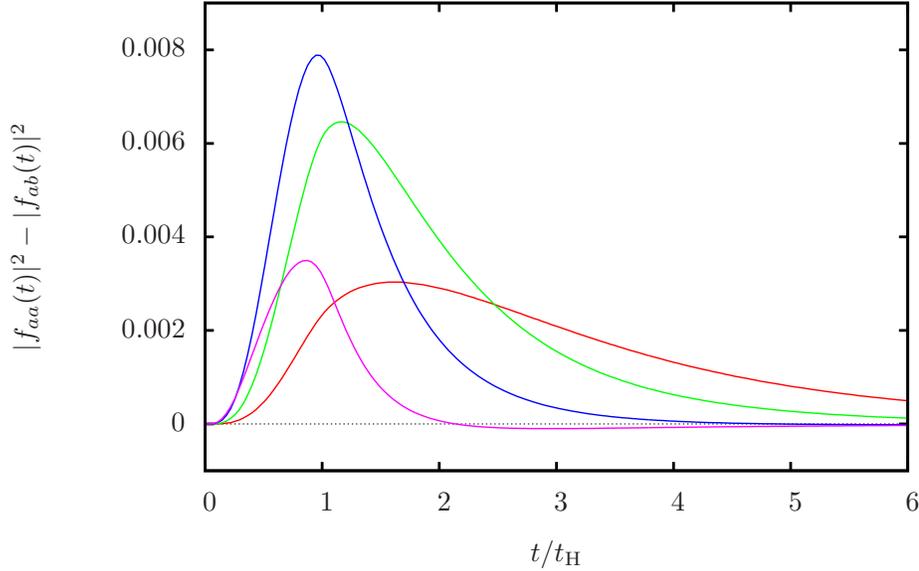}
\caption{Comparison between fidelity decay, measured in reflection 
$|f_{aa}(t)|^2$ and in transmission $|f_{ab}(t)|^2$. Different colours 
correspond to different perturbation strengths, $\delta_1 = 0.07$ (red line), 
$0.14$ (green line), $0.28$ (blue line), and $0.56$ (pink line).}
\label{Fig3}\end{figure}

In the perturbative result,~\eref{BRW:Eq2}, it makes no difference
whether the measurement is performed as a transmission measurement with two 
measurement antennas or as a reflection measurement with only one. As we have 
seen from~\eref{F:fab} and~\eref{F:faa}, for the exact analytical
result this is no longer true. In \fref{Fig3}, we compare both cases for
four different perturbation strengths. The results show that the difference 
$|f_{aa}(t)|^2 - |f_{ab}(t)|^2$ is clearly present, but usually quite small.
For very large ($\delta_1= 0.56$, narrow peak at $t \lesssim t_H$) as well 
as for very small perturbation strengths, it seems that 
the difference tends to disappear. This is consistent with our
treatment of the perturbative case in Sec.~\ref{AL}.

\section{\label{A} Perturbative regime}

In the perturbative regime, the fidelity amplitude of \eref{fidstart} can be
calculated using first order time-independent perturbation 
theory~\cite{CerTom02}. For $\{ |j\ra \}$ denoting the eigenbasis of $H_\beta$ 
and $V= H_\alpha - H_\beta$, we then find 
\begin{equation}
\fl f(t) \approx \sum_j \la\alpha|j\ra\; \rme^{2\pi\rmi E_j(\beta)\, t}\;
   \rme^{-2\pi\rmi [E_j(\beta) + \la j|V|j\ra ]\, t}\; \la j|\alpha\ra
 = \sum_j |\la j|\alpha\ra|^2\; \rme^{-2\pi\rmi \la j|V|j\ra\, t} \; ,
\end{equation}
where $|\alpha\ra$ denotes some initial state, and $\{ E_j(\beta) \}$ denote
the eigenvalues of $H_\beta$. This expression shows that in the perturbative
regime the fidelity decay depends on the product between time and perturbation
strength. The perturbative result becomes exact only in the limit of vanishing
perturbation strength. To yield a finite value for the fidelity amplitude,
time must then tend to infinity such that the product between perturbation 
strength and time remains constant. We therefore define the perturbative regime
as the limit
\begin{equation}
t\to\infty \; , \quad \forall j\; :\; T_j\to 0\; ,
\label{A:pertregdef}\end{equation}
such that $t\, \sum T_j$ and $t^2\, \sum_j T_j^2$ remain both finite. In what
follows, we calculate $f_{ab}(t)$ and $f_{aa}(t)$ in that limit, starting from
the exact analytical expressions~\eref{F:fab} and~\eref{F:faa}, via an 
asymptotic expansion of the respective integrals. This is done in two steps.

\subsection{Step one}

Here, we will demonstrate that 
\begin{eqnarray}
\fl \la\ldots\ra_I \sim \la\ldots\ra_A \; , \quad{\rm with}\quad 
\la\ldots\ra_A = \int_{t-1}^t\rmd r\int_0^{\sqrt{r}}\rmd u\; 
   \frac{(t-r)(r+1-t)}{(2u+1)(t^2-r^2+x^2)^2} \; , 
\label{A:StepOne}\end{eqnarray}
where the ellipsis above may be replaced by either $Z\, P_0$ or $4t^2\, Z$.
Here and in what follows, the symbol $\sim$ denotes the perturbative limit we
are interested in.

For the first case, our claim follows from 
\begin{equation}
\int_{t-1}^t\rmd r\int_{\sqrt{r}}^r\rmd u\; 
   \frac{(t-r)(r+1-t)\; P_0\; Z}{(2u+1) (t^2 - r^2 + x^2)^2} \sim 0 \; .
\label{MS:asym1}\end{equation}
Since $P_0 > 0 $ and $0 < Z < 1$ in the 
whole region of integration, it is sufficient to show that~\eref{MS:asym1}
holds for $Z=1$. Furthermore, since we can maximize $(t-r)(r+1-t)$ in the 
interval $t-1 < r < t$ by $1/4$, it is sufficient to prove that 
\begin{equation}
\max_{t-1 < r < t} \int_{\sqrt{r}}^r\rmd u\; 
   \frac{r^2 + (2r+1)t - t^2 - x^2}{(2u+1) (t^2 - r^2 + x^2)^2} \sim 0 \; ,
\end{equation}
where we have used that $P_0= 2\, (r^2 + (2r+1)t - t^2 - x^2)$. Denoting this
integral with $J$, we realize that
\begin{equation}
J < \int_{\sqrt{r}}^r\rmd u\; 
   \frac{r^2 + (2r+1)t - t^2}{(2u+1) (t^2 - r^2 + x^2)^2}
\end{equation}
Then, because $r^2 + (2r+1)t - t^2 = t + 2r^2 - (r^2 -2rt + t^2)$,
\begin{equation}
J < \int_{\sqrt{r}}^r\rmd u\; \frac{2t^2 + t}{(2u+1) x^4} 
 = (2t^2 + t)\int_{\sqrt{r}}^r\rmd u\; \frac{2u+1}{(2r+1)^2\, u^4}
\label{A:Jsmaller}\end{equation}
Evaluating the last integral we finally obtain:
\begin{equation}
J < \frac{t\, (2t+1)}{(2r+1)^2}
   \left( \frac{3\sqrt{r} + 1}{3r^{3/2}} - \frac{3r+1}{3r^3}\right) = 0\; ,
\end{equation}
which completes the proof. For the second case, we replace $P_0$ with $2t^2$ 
one arrives at the same result, which may be seen from~\eref{A:Jsmaller}.

\subsection{Step two}

According to~\eref{A:StepOne} the perturbative limit only requires 
integration of $u$ up to $u=\sqrt{r}$. This implies that
\begin{equation}
Z= \prod_j \frac{1 - T_j (t-r)}{\sqrt{1 + 2 T_j\, r + T_j^2\, x^2}}
 \sim \prod_j \frac{1}{\sqrt{1 + 2 T_j\, t}} \; ,
\end{equation}
since $(t-r)$ is of order one, $t-1 < r < t$, and 
\begin{equation}
T_j^2\, x^2 = T_j^2\, u^2\; \frac{2r+1}{2u+1} 
  < T_j^2\, r\; \frac{2r+1}{2\sqrt{r} +1} \sim 0 \; .
\end{equation}
Therefore, we obtain for $f_{ab}(t)= \la Z\, P_0\ra_I$:
\begin{equation}
f_{ab}(t) \sim \prod_j \frac{1}{\sqrt{1 + 2 T_j\, t}}
  \la P_0\ra_A \sim \prod_j \frac{1}{\sqrt{1 + 2 T_j\, t}} \; .
\end{equation}
This simply follows from the fact that $1 = \la P_0\ra_I \sim \la P_0\ra_A$.
For the scattering fidelity in a reflection measurement, we obtain
\begin{equation}
f_{aa}(t) \sim \frac{4t^2\, \la Z\ra_I + 2\, \la Z\, P_0\ra_I}{3} \; ,
\end{equation}
since $b_2(t)$ in~\eref{F:faa} tends to zero for large times. Here, it 
only remains to treat the first term in the nominator:
\begin{equation}
4t^2\, \la Z\ra_I \sim \prod_j \frac{1}{\sqrt{1 + 2 T_j\, t}}\; 4t^2\; 
  \la 1\ra_A \sim \prod_j \frac{1}{\sqrt{1 + 2 T_j\, t}} \; ,
\end{equation}
since $1= 4t^2\, \la 1\ra_I \sim 4t^2\; \la 1\ra_A$. Thus, in the perturbative
regime, we obtain the same result no matter whether we perform a transmission
or a reflection measurement:
\begin{equation}
f_{aa}(t) \sim f_{ab}(t) \sim 
f_{\rm pert}(t) = \prod_j \frac{1}{\sqrt{1 + 2 T_j\, t}} \; .
\label{A:fpert}\end{equation}

\subsection{Global perturbation}

Global perturbations are discussed in detail in Sec.~\ref{LG}, 
where~\eref{LG:Teffkapp} relates the effective transmission coefficients with
the perturbation strengths $\delta_j$. Taking also~\eref{LG:kapisums} into 
account, we may write for $f(t)$ up to second order in the perturbation 
strength:
\begin{equation}
\fl \ln\, f_{\rm pert}(t) \sim -\;\frac{1}{2}\sum_j \ln\big [\, 
   1+4\pi\rmi\, \delta_j\, (1-\rmi\pi\, \delta_j)\, t\, \big ]
 \sim -\sum_j \big [\, 2\pi\rmi\, \delta_j\, t + 2\pi^2\delta_j^2\, (2t^2 + t)
   \, \big ] \; .
\end{equation}
Since we are working in the perturbative regime, where $t$ goes as fast to
infinity as the $\delta_j$ go to zero, terms containing $\delta_j^2\, t$ can be
neglected. Finally, we obtain
\begin{equation}
\ln\, f_{\rm pert}(t) \sim \rme^{-2\pi\rmi\, \delta_{\rm s}\, t 
   - 4\pi^2\lambda^2\, t^2} \; .
\end{equation}

\subsection{\label{AL} Moving scatterer}

Inserting the effective transmission coefficients from~\eref{LL:T1T2}
describing a moving scatterer into~\eref{A:fpert}, we find
\begin{equation}
f_{\rm pert}(t) = 
   \left|1 - \frac{4\rmi\, \delta_1\, t}{1 - \rmi\, \delta_1}\right|^{-1} \; .
\end{equation}
In the perturbative limit considered here, means that $T_1\to 0$, $t\to\infty$ 
such that $T_1\, t$ remains constant. This implies however that also
$\delta_1 \to 0$ such that $\delta_1\, t$ remains constant. Therefore
\begin{equation}
f_{\rm pert}(t) \sim \left|1 - 4\rmi\, \delta_1\, t\right|^{-1}
 = \frac{1}{\sqrt{1 + (4\delta_1\, t)^2 }} \; .
\label{A:movscat}\end{equation}

\section{\label{P} Comparison with parametric level correlations}

For any quantum mechanical model of the form
\begin{equation}
H(\lambda) = H_0 + \lambda\; V \; ,
\end{equation}
we may consider the level dynamics obtained from plotting the 
eigenvalues of $H(\lambda)$ as functions of $\lambda$. For convenience, we 
assume here again that for any value of $\lambda$, the average level spacing is 
one. In a typical random matrix model, one would eventually choose $V$ from
a Gaussian random ensemble with non-diagonal elements of unit variance.

The parametric level correlations $X(\lambda,r)$ describe the probability to
find two eigenvalues, one of $H(0)$ and the other one of $H(\lambda)$ at a 
distance $r$. The quantity to be compared to the fidelity amplitude is the 
Fourier transform of the parametric level correlations~\cite{Koh08}:
\begin{equation}
K(\lambda,t) = \int\rmd r\; \rme^{2\pi\rmi\, rt}\; \big [\, X(\lambda,r) - 1
   \, \big ] \; .
\end{equation}
Note that for $\lambda\to 0$, this quantity converges to the complement of the 
two-point form factor: $K(0,t)= 1 - b_2(t)$~\cite{GMW98,Meh2004}.%
\footnote{The definition of $K(\lambda,t)$ in~\cite{Koh08} uses 
the wrong sign, while the final expressions for $K_1(\lambda,t)$ misses the 
variable $v$ in the nominator of the integrand.}
The relation discussed in~\cite{Koh08} is a relation between the 
parametric level correlations on the one hand, and the fidelity amplitude 
$f_\lambda(t) = f(\lambda,t)$ on the other. It may be expressed as
\begin{equation}
f(\lambda,t) = \frac{-\beta}{4\pi^2 t^2}\; 
  \frac{\partial}{\partial\, (\lambda^2)}\; K(\lambda,t) \; ,
\label{FidParaCor}\end{equation}
with $\beta$ being the Dyson parameter~\cite{Dys62c} which is one in our case.
We consider systems with an anti-unitary symmetry such as time reversal 
invariance.

\subsection{Global perturbation}

From~\cite{MaSmSi03} we find
\begin{equation}
\fl X(\lambda,r) = 1 + {\rm Re}\int\!\!\!\int_1^\infty\rmd\lambda_1\rmd\lambda_2
  \int_{-1}^1\rmd\mu'\; \frac{(\lambda_1\lambda_2 - \mu')^2\, 
    (1-\mu'{}^2)\, \rme^{\rmi\pi\, r_+\, (\lambda_1\lambda_2 - \mu')}\,
    \rme^{-\pi^2 \lambda^2\, P_0}}
  {(1 + 2\lambda_1\lambda_2\, \mu' -\lambda_1^2 -\lambda_2^2 -\mu'{}^2)^2}\; ,
\end{equation}
where 
$2P_0 = 1+ 2\lambda_1^2\lambda_2^2 -\lambda_1^2 -\lambda_2^2 - \mu'{}^2$. We 
will see below, that this quantity is precisely the same as $P_0$ defined in 
the previous section, in~\eref{JPlim}. The first substitution,
$\mu'\to \mu= (\lambda_1\lambda_2 -\mu')/2$, yields
\begin{equation}
\fl X(\lambda,r) = 1 + 2\; {\rm Re}\int\!\!\!\int_1^\infty\rmd\lambda_1\rmd\lambda_2
  \int_{(\lambda_1\lambda_2-1)/2}^{(\lambda_1\lambda_2+1)/2}\rmd\mu\;
  \frac{4\mu^2\, (1-\mu'{}^2)\, \rme^{2\rmi\pi\, r_+\, \mu}\,
     \rme^{-\pi^2 \lambda^2\, P_0}}
  {(1 + 2\lambda_1\lambda_2\, \mu' -\lambda_1^2 -\lambda_2^2 -\mu'{}^2)^2}\; .
\label{XcorrelInt}\end{equation}
In order to shorten the expressions, we keep writing $\mu'$ which must be 
understood as being a function of $\mu$. Now, we can switch to the Fourier 
transform, which turns the Fourier factors in delta functions:
\begin{equation}
\fl K(\lambda,t) = \int\!\!\!\int_1^\infty\rmd\lambda_1\rmd\lambda_2
  \int_{(\lambda_1\lambda_2-1)/2}^{(\lambda_1\lambda_2+1)/2}\rmd\mu
  \frac{\big [\, \delta(t+\mu) + \delta(t-\mu)\, \big ]\; 
  4\mu^2\, (1-\mu'{}^2)\, \rme^{-\pi^2 \lambda^2\, P_0}}
  {(1 + 2\lambda_1\lambda_2\, \mu' -\lambda_1^2 -\lambda_2^2 -\mu'{}^2)^2}\; .
\label{KcorrelInt}\end{equation}
This shows that the function $K(\lambda,t)$ is symmetric in time. In what 
follows, we thus assume $t>0$. The remaining delta function already allows to 
eliminate the $\mu$-integration. However, before actually doing so, we perform 
a variable transformation on the $\lambda_1,\lambda_2$ integrals:
\begin{equation}
(\lambda_1,\lambda_2) \to (r',x')\; , \qquad r'= \lambda_1\lambda_2\; , \quad
x'= \lambda_2 - \lambda_1 
\end{equation}
The Jacobian of this transformation is simply 
$J= (\lambda_1 + \lambda_2)^{-1} = (x'{}^2 + 4r')^{-1/2}$. Therefore,
\begin{eqnarray}
\fl K(\lambda,t) = \int_1^\infty\rmd r'\int_{1-r'}^{r'-1}
  \frac{\rmd x'}{\sqrt{x'{}^2 + 4r'}}\int_{(r'-1)/2}^{(r'+1)/2}\rmd\mu\nonumber\\
 \frac{4\mu^2\, (1-r'+2\mu)(1+r'-2\mu)\, \delta(t-\mu)\,
   \rme^{-\pi^2 \lambda^2\, P_0}}
   {[1+2r'(r'-2\mu) - x'{}^2 - 2r' - (r'-2\mu)^2 ]^2}
\end{eqnarray}
For the $\mu$-integral not to yield zero, it must hold that 
$(r'-1)/2 < t < (r'+1)/2$. This modifies the limits of the $r'$-integral as
follows:
\begin{equation}
\fl K(\lambda,t) = 4t^2\int_{\max(1,2t-1)}^{2t+1}\!\!\!\!\!\rmd r'\!\!\!
  \int_{1-r'}^{r'-1}\!\!\!
  \frac{\rmd x'}{\sqrt{x'{}^2 + 4r'}}\;
   \frac{(1-r'+2t)(1+r'-2t)\, \rme^{-\pi^2 \lambda^2\, P_0}}
   {[1+2r'(r'-2t) - x'{}^2 - 2r' - (r'-2t)^2 ]^2}
\end{equation}
Further substitutions: $r'= 2r+1$ and $x'= 2x$ and the fact that the integrand
is a symmetric function of $x$, yield
\begin{equation}
\fl K(\lambda,t) = 4\, t^2\int_{\max(0,t-1)}^t\rmd r\int_0^r
  \frac{\rmd x\, (t-r)\, (r+1-t)\, \rme^{-\pi^2 \lambda^2\, P_0}}
   {\sqrt{x^2 + 2r+1}\, (t^2 - r^2 + x^2)^2}
 = 4\, t^2\; \lla \rme^{-\pi^2 \lambda^2\, P_0}\rra_I\; .
\label{Kfinal}\end{equation}
Comparing to~\eref{FidGlobal}, it is now easily checked that $K(\lambda,t)$ as 
defined here fulfils the fidelity amplitude -- parametric form factor 
relation~\eref{FidParaCor}.

\subsection{General perturbation}

In~\cite{MaSmSi03} it is shown that parametric level correlations can be
calculated for arbitrary perturbations. According to this reference, the term 
describing the global perturbation $\sigma_{\rm glob} = \pi^2\lambda^2\, P_0$
must be replaced by $\sigma = \sigma_{\rm glob} + \sigma_{\rm loc}$, where
\begin{equation}
\sigma_{\rm loc}(\lambda_1,\lambda_2,\mu') = \frac{1}{2}\sum_j \ln\!\left[
   \frac{1+2\rmi\, \kappa'_j\, \lambda_1\lambda_2 - \kappa'_j{}^2\, 
   (\lambda_1^2 + \lambda_2^2 -1)}{(1+\rmi\, \kappa'_j\, \mu')^2} \right] \; .
\end{equation}
Note that the discussion in Sec.~\ref{LG} shows that the additional global 
perturbation could always be incorporated into $\sigma_{\rm loc}$, via a 
large number of additional channels with infinitesimal perturbations. However,
in order to establish the desired relation between fidelity decay and the 
parametric level correlations, it is important to have the parameter $\lambda$
describing the global perturbation at hand.

Now, we should go through the calculation of $K(\lambda,t)$ again, replacing
$\sigma_{\rm glob}$ in~\eref{XcorrelInt} with the more general expression 
$\sigma$. As a consequence, the integrand is no longer real, which affects
\eref{KcorrelInt}. While the delta function $\delta(t-\mu)$ is multiplied with 
the same term as before, the second delta function $\delta(t+\mu)$ is now 
multiplied with its complex conjugate. Therefore $K(\lambda,t)$ is no longer
symmetric. Instead $K(\lambda,t) = K(\lambda,-t)^*$, which nevertheless allows 
to continue the calculation without changes for $t > 0$. Only at~\eref{Kfinal} 
we need to express $\sigma_{\rm loc}$ in the current integration variables. 
That results in
\begin{equation}
\exp\big [ -\sigma_{\rm loc}(r,x,t)\, \big] 
 = \prod_j \frac{1- T_j\, (t-r)}{\sqrt{1 + 2 T_j\, r + T_j^2\, x^2}} \; , 
\end{equation}
where $T_j = 2\rmi \kappa_j'/(1 + \rmi\, \kappa_j')$, just as 
in~\eref{LG:Teffkapp}. Inserting this expression into~\eref{Kfinal} and
comparing to the general result~\eref{F:fab} for the decay of the fidelity 
amplitude ($a\ne b$) we find that the following relation holds:
\begin{equation}
f_{ab}(\{ \kappa_j'\},t) = \frac{-1}{4\pi^2\, t^2}\; 
   \left. \frac{\partial}{\partial (\lambda^2)}\, K(\{ \kappa_j'\}, \lambda,t)
   \right|_{\lambda = 0} \; .
\end{equation}
The fact that any global perturbation may always be modelled with a large
number of additional terms in $\sigma_{\rm loc}$, allows for a final slight
generalization:
\begin{equation}
f_{ab}(\{ \kappa_j'\},\lambda_0,t) = \frac{-1}{4\pi^2\, t^2}\; 
   \left. \frac{\partial}{\partial (\lambda^2)}\, K(\{ \kappa_j'\}, \lambda,t)
   \right|_{\lambda = \lambda_0} \; .
\label{FinalEq}\end{equation}
This relation constitutes our second important result. In practice, this 
relation means that one can obtain the fidelity amplitude, by measuring the 
change of the parametric level correlations under the increment of a global 
perturbation.

\section{\label{C} Conclusions}

In this paper, we have used the results of~\cite{KKSGSS10} to derive an
exact analytical expression for the fidelity decay in a closed 
chaotic/diffusive wave system, under arbitrary Hermitian perturbations. For 
illustration, we used that result to re-derive the known formula for the 
fidelity decay in the case of a global perturbation~\cite{StoSch04b,StoSch05}. 
In a second application, we calculated the fidelity amplitude for a moving
S-wave scatterer, and checked that it describes corresponding experimental
results reported in~\cite{Hoe08} well. Finally, we generalized a relation 
between the fidelity amplitude and parametric level correlations introduced
in~\cite{Koh08} to arbitrary perturbations. 

In the present work, we restricted ourselves to matrix ensembles based on the 
Gaussian orthogonal ensemble (GOE). For the comparison with experimental data,
this is the most important case. However, our results can also be translated
to the Gaussian unitary ensemble (GUE), using the analog of the VWZ-formula 
published in~\cite{SFS06}. For the Gaussian symplectic ensemble (GSE), the 
corresponding analytical expressions for the parametric level correlations and 
the correlations between scattering matrix elements are unfortunately not yet 
available, but we would still expect a similar relation to hold.

It would be interesting to perform an experiment similar to the one analyzed
in~\cite{BKS99,Hoe08}, in order to verify our results with higher accuracy 
and for larger perturbation strengths. Particularly interesting would be the 
regime, where the perturbation strength depends in a non-linear way on the 
displacement of the scatterer, see~\eref{BRW:Eq1}. If the microwave 
experiment would allow to measure fidelity decay and parametric level 
correlations at the same time, one could test the applicability of the 
relation~\eref{FinalEq} between both quantities practice. Finally, one may
intend to generalize~\eref{FinalEq} further to scattering systems and 
non-Hermitian perturbations (\eg coupling fidelity).

\ack
We are very grateful to R. H\" ohmann, U. Kuhl, and H.-J. St\" ockmann for 
providing us with the experimental data for the fidelity decay in the case of a 
moving scatterer. We thank H.-J. St\" ockmann, H. Kohler, U. Kuhl, and T. H. 
Seligman for useful discussions, and acknowledge the hospitality of the Centro 
Internacional de Ciencias, UNAM, where some of these discussions took place. 
Finally, we acknowledge financial support from CONACyT through the grant 
CB-2009/129309.


\section*{References}

\bibliographystyle{iopart-num}
\bibliography{/home/gorin/Bib/JabRef}

\end{document}